\documentclass[preprint,12pt]{elsarticle}
\usepackage{graphicx}
\usepackage{epsfig}
\usepackage{amssymb}
\usepackage{amsmath}
\begin{document}
\begin{frontmatter}
\title{Systematic Properties of the Tsallis Distribution:\\
 Energy Dependence of Parameters in High-Energy $p-p$ Collisions}
\author{J. Cleymans$^1$, G.I. Lykasov$^2$, A.S.  Parvan$^{2,3}$, A.S. Sorin$^2$,
 O.V.~Teryaev$^2$, D.~Worku$^1$}
\address{$^1$ UCT-CERN Research Centre and Department of Physics,
University of Cape Town, Rondebosch 7701, Cape, Republic of South Africa\\ 
$^2$ JINR, Dubna, 141980, Moscow region, Russia\\
$^3$ Institute of Applied Physics, Moldova Academy of Sciences,
MD-2028 Chisinau, Republic of Moldova}
\begin{abstract}
Changes in the transverse momentum distributions with beam energy
are studied using the Tsallis distribution as a parameterization.
The dependence of the Tsallis parameters $q$, $T$ and  the volume  
 are determined as a function of beam energy.
The Tsallis parameter $q$ shows a weak but clear increase with beam energy
with the highest value being approximately 1.15.
The Tsallis temperature and volume are consistent with being independent 
of beam energy within experimental uncertainties.
\end{abstract}
%

\end{frontmatter}


%
\section{\label{sec:Introduction}Introduction}
There exists a rich and wide variety of distributions covering a 
large range of applications~\cite{newman,clauset,levy}.
Those having a power law behaviour have attracted considerable attention in physics in 
recent years but there is a a long history 
in other fields such as biology and economics~\cite{mitzenmacher}.

In high energy physics power law distributions have been applied 
in~\cite{STAR,PHENIX,ALICE,ATLAS,CMS} to the description of transverse momenta  
of secondary particles produced in $p-p$ collisions. 
Indeed the available 
range of transverse momenta  has expanded
considerably  with the advent of the Large Hadron Collider (LHC). 
Collider  energies of 8 TeV are now available in $p-p$ collisions and 
transverse momenta of hundreds of GeV are now  common.
Applications of the Tsallis 
distribution to high energy $e^+ - e^-$ annihilation has been considered previously
 in~\cite{bediaga}. A recent review of
power laws in elementary and heavy-ion collisions 
can be found in~\cite{wilkreview}.
The Tsallis distribution has the advantage of being connected, via the entropy, to thermodynamics
which is not the case for other power-law distributions~\cite{hagedorn,michael}.
In this paper the focus is on the energy dependence of the Tsallis parameters 
describing the transverse momenta at mid-rapidity in high 
energy $p-p$ collisions.   
The rapidity distributions contain more dynamics, e.g. strong longitudinal expansion, and
will not be considered in this paper which is limited to mid-rapidity.
%
\section{Choice of Distribution}
In the analysis of the new data, a Tsallis-like distribution 
gives excellent fits to the transverse momentum 
distributions  as shown by the  
 STAR~\cite{STAR} and PHENIX~\cite{PHENIX} collaborations at RHIC and by the  
ALICE~\cite{ALICE}, ATLAS~\cite{ATLAS} and CMS~\cite{CMS} collaborations at 
the LHC.
In this paper we review the parameterization used by these groups and propose a slightly different one
which leads to a more consistent interpretation and has the bonus of being  thermodynamically consistent.

In the framework of Tsallis 
statistics~\cite{tsallis1,tsallis2,miller,worku1,worku2,parvan1,parvan2}
the particle number, $N$,  the energy density $\epsilon$ and the pressure $P$ 
are given by corresponding
integrals over the  Tsallis distribution  
%
\begin{eqnarray}
N &=& gV\int\frac{d^3p}{(2\pi)^3}
\left[ 1 + (q-1) \frac{E-\mu}{T}\right]^{-\frac{q}{q-1}} ,\label{number} \\
\epsilon &=& g\int\frac{d^3p}{(2\pi)^3}E
\left[ 1 + (q-1) \frac{E-\mu}{T}\right]^{-\frac{q}{q-1}} ,\label{epsilon}\\
P &=& g\int\frac{d^3p}{(2\pi)^3}\frac{p^2}{3E}
\left[ 1 + (q-1) \frac{E-\mu}{T}\right]^{-\frac{q}{q-1}}\label{pressure} ,\\
\epsilon +P &=& Ts + \mu n \label{entropy}.
\end{eqnarray}
where $T$ and $\mu$ are the temperature and the chemical potential,
$V$ is the volume and  $g$ is the degeneracy factor.  
These expressions are  thermodynamically consistent, 
e.g. it can be easily shown~\cite{worku2} that consistency relations
of the type
\begin{equation}
 N = V\left.\frac{\partial P}{\partial \mu}\right|_{T},\label{consistency}
\end{equation}
are indeed satisfied. 
The parameter $q$ might be treated
as a sort of an anomalous dimension, while the parameter
$T$ characterizes a soft dimensionfull scale.

Note the extra power of $q$ compared to other forms of the Tsallis 
distribution, this is necessary in order to have  thermodynamic consistency.
The motivation for using the above expressions has been presented 
in detail in~\cite{worku2} and will not repeated here.
Other proposals have been made in the literature, 
see e.g.~\cite{lavagno1,lavagno2,lavagno3} and using a mean field approach~\cite{lavagno4}.
The particular form of the Tsallis distribution presented above has been referred to as Tsallis-B
distribution in~\cite{worku2}.
Following from~\eqref{number}, the momentum distribution is given by,
\begin{equation}
\frac{d^{3}N}{d^3p} = 
\frac{gV}{(2\pi)^3}
\left[1+(q-1)\frac{E -\mu}{T}\right]^{-q/(q-1)},
\label{tsallismu}
\end{equation}
or, expressed in terms of transverse momentum, $p_T$,  
the transverse mass, $m_T \equiv \sqrt{p_T^2+ m ^2}$, and the rapidity  $y$  
\begin{equation}
\frac{d^{2}N}{dp_T~dy} = 
gV\frac{p_Tm_T\cosh y}{(2\pi)^2}
\left[1+(q-1)\frac{m_T\cosh y -\mu}{T}\right]^{-q/(q-1)} .
\label{tsallismu1}
\end{equation}
At mid-rapidity, $y = 0$, and for zero chemical potential, as is relevant at 
the LHC, this reduces to 
\begin{equation}
\left.\frac{d^{2}N}{dp_T~dy}\right|_{y=0} = 
gV\frac{p_Tm_T}{(2\pi)^2}
\left[1+(q-1)\frac{m_T}{T}\right]^{-q/(q-1)}.
\label{tsallisfit1}
\end{equation}
In the limit where the parameter $q$ goes to 1 it is well-known that this reduces 
the standard Boltzmann distribution:
\begin{equation}
\lim_{q\rightarrow 1}\frac{d^{2}N}{dp_T~dy} = 
gV\frac{p_Tm_T\cosh y}{(2\pi)^2}
\exp\left(-\frac{m_T\cosh y -\mu}{T}\right).
\label{boltzmann}
\end{equation}
The parameterization given in Eq.~\eqref{tsallismu1} is close to
the one used e.g. by the  ALICE and other
collaborations~\cite{STAR,PHENIX,ALICE,ATLAS,CMS}: 
\begin{equation}
  \frac{d^2N}{dp_T\,dy} = p_T \frac{dN}
  {dy} \frac{(n-1)(n-2)}{nC(nC + m_{0} (n-2))}
 \left[ 1 + \frac{m_T - m_{0}}{nC} \right]^{-n}  ,
\label{alice}
\end{equation}
where $n$ and $C$ are fit parameters.
This formula can also be understood as an interpolation between low 
transverse momenta exponential fall-off and high momenta QCD power 
law behaviour~\cite{hagedorn,michael}.
 This corresponds to substituting 
\begin{equation}
n\rightarrow \frac{q}{q-1}   ,
\label{n}
\end{equation}
and 
\begin{equation}
nC  \rightarrow \frac{T+m_0(q-1)}{q-1}  .
\label{nC}
\end{equation}
After  this substitution Eq.~\eqref{alice} becomes
\begin{eqnarray}
  \frac{d^2N}{dp_T\,dy} =&& p_{T} \frac{{\rm d}N}
  {{\rm d}y} \frac{(n-1)(n-2)}{nC(nC + m_{0} (n-2))}\nonumber\\ 
&&\left[\frac{T}{T+m_0(q-1)}\right]^{-q/(q-1)}\nonumber\\
&&\left[ 1 + (q-1)\frac{m_T}{T} \right]^{-q/(q-1)}  .
\label{alice2}
\end{eqnarray}
At mid-rapidity $y=0$ and zero chemical potential,
this has the same dependence on the 
transverse momentum as~\eqref{tsallisfit1} 
apart from an additional  factor $m_T$ on the right-hand side of~\eqref{tsallismu1}.
However, the inclusion of the rest mass in the substitution Eq.~\eqref{nC}
is not in agreement with the Tsallis distribution as it breaks 
$m_T$ scaling which is present in Eq.~\eqref{tsallismu1}
 but not in Eq.~\eqref{alice}. 
The inclusion of the factor $m_T$ 
leads to a more consistent interpretation of the variables $q$ and $T$.
\\
It is to be noted the variables $(T,V,q,\mu)$ in the 
distribution function Eq.~\eqref{tsallismu} have a redundancy for $\mu \neq 0$
and are not independent.  Indeed, let $T=T_{0}$ and $V=V_{0}$ at $\mu=0$ 
and fixed values of $q$. 
Comparing Eq.~\eqref{tsallismu} written for $\mu=0$ and the same equation 
written for finite values of $\mu$, we obtain
\begin{eqnarray}
  T_0 &=& T \left[1-(q-1) \frac{\mu}{T} \right], \qquad  \mu\leq\frac{T}{q-1},
\label{T0}  \\ 
  V_0 &=& V  \left[1-(q-1) \frac{\mu}{T} \right]^{\frac{q}{1-q}}
\label{V0}.
\end{eqnarray}
Therefore, the variables $T$ and $V$ are functions of $\mu$ at 
fixed values of $q$ and they can be calculated if the 
parameters $(T_{0},V_{0})$ and $q$ are known. This redundancy is not present when $\mu = 0$, which is the case relevant for the LHC due
to the particle-antiparticle symmetry.

A very good description of transverse momenta distributions at RHIC has been
obtained in Refs~\cite{coalesce1,coalesce2} on the basis of a coalescence model 
where the Tsallis distribution is used for quarks.  In this paper we use the Tsallis distribution for hadrons, not for quarks.\\

A Tsallis fit has also been considered in Ref.~\cite{wong} but a different power law in the Tsallis function was considered by these authors.\\
Interesting results were obtained in 
Refs.~\cite{deppman,deppman1,deppman2,deppman3}
where spectra for identified particles were analyzed and the resulting
values for the parameters $q$ and $T$ were considered. These authors did not 
consider the energy dependence which is the main focus of the present paper.\\
The transverse momentum distribution in connection with the multiplicity in
different events was consider in Ref.~\cite{urmossy}.\\
The energy dependence of the transverse momentum spectra has been studied 
on the basis of gluon collisions in Ref.~\cite{gena}.\\
\section{Energy Dependence of Transverse Momentum Distributions} 
%
The energy dependence in $p-p$ collisions can be determined by 
studying data at beam energies
of 0.54~\cite{UA1}, 0.9, 2.36  and 7 TeV~\cite{ATLAS,CMS}. 
These involve distributions summed over charged particles.
The fits were performed using a sum of three Tsallis distributions, the 
first one for $\pi^+$, the
second one for $K^+$ and the third one for protons $p$. The relative 
weights between these were 
simply determined by the corresponding degeneracy factors, i.e. 1 for for $\pi^+$ and $K^+$
and 2 for  protons. 
The fit was taken at mid-rapidity and for $\mu = 0$ using the following expression:
\begin{equation}\label{tsallisfit}
   \left.  \frac{1}{2\pi p_{T}} \frac{d^{2}N(\mathrm{charged \
particles})}{dp_{T}dy}\right|_{y=0} = \frac{2V}{(2\pi)^{3}}
\sum\limits_{i=1}^{3} g_{i} m_{T,i}
\left[1+(q-1)\frac{m_{T,i}}{T}\right]^{-\frac{q}{q-1}},
\end{equation}
where $i=(\pi^{+},K^{+},p)$  and
$g_{\pi^{+}}=1$, $g_{K^{+}}=1$ and $g_{p}=2$. The factor $2$ in front
of the right hand side of this equation takes into account the
contribution of the antiparticles $(\pi^{-},K^{-},\bar{p})$. 
The resulting values of the parameters  are shown in Table 1. 

The Tsallis parameter $q$ has a tendency to increase slowly
but clearly  with increasing  energy.

\begin{figure}
\begin{center}
\includegraphics[width=\textwidth,height=12cm]{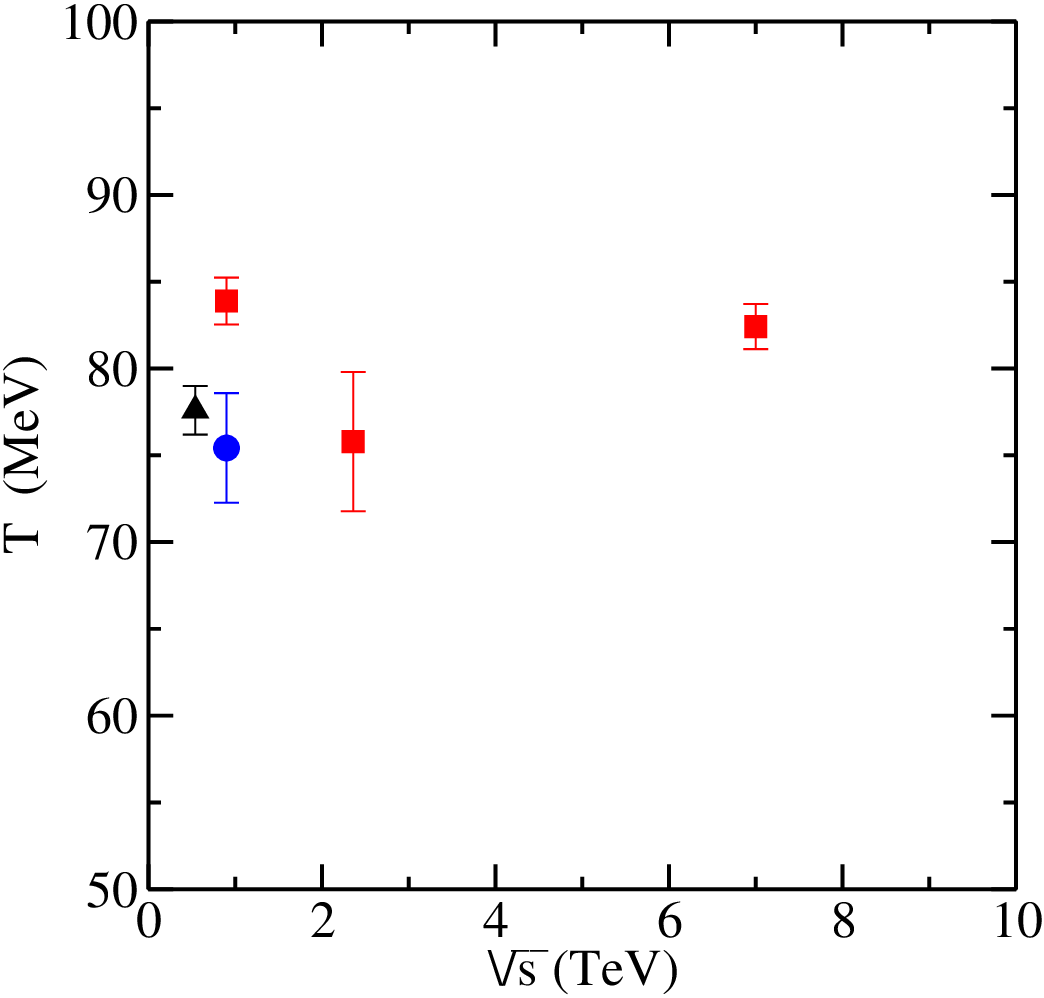}
\caption{(Color online) 
Energy dependence of the Tsallis temperature $T$ appearing in the Tsallis distribution.
Square points are from the ATLAS collaboration~\cite{ATLAS}, 
the round point is from the ALICE collaboration~\cite{ALICE} and the triangle point is from the UA1 collaboration~\cite{UA1}.
}
\label{tsallis_T_v3}
\end{center}
\end{figure}

\begin{figure}
\begin{center}
\includegraphics[width=\textwidth,height=12cm]{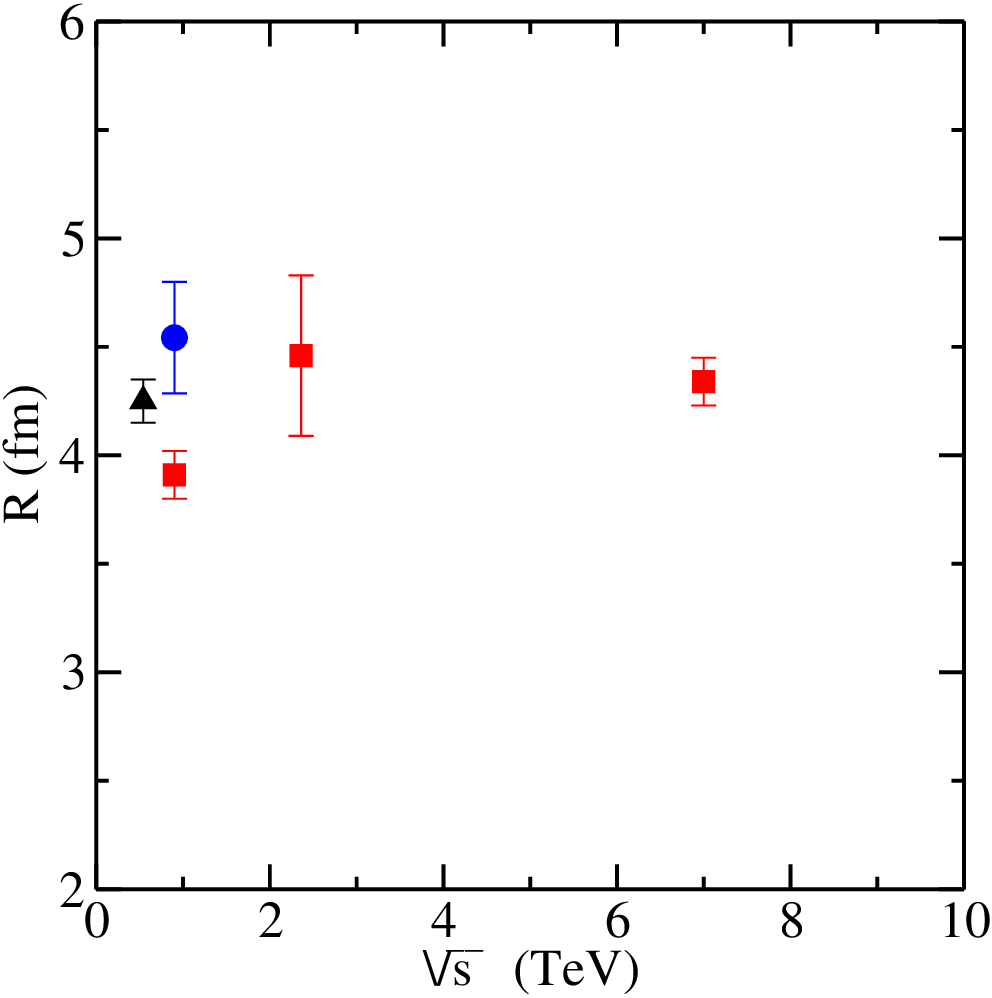}
\caption{(Color online)
Energy dependence of the Tsallis radius $R$ appearing in the 
volume factor.
Square points are from the ATLAS collaboration~\cite{ATLAS}, 
the round point is from the ALICE collaboration~\cite{ALICE} and the triangle point is from the UA1 collaboration~\cite{UA1}.
}
\label{tsallis_R_v3}
\end{center}
\end{figure}

The radius is listed instead of the volume $V$  in Eq.~\eqref{tsallisfit} and is defined as
$R \equiv \left[V \frac{3}{4\pi}\right]^{1/3}$.
Note that the volume can contain dynamical factors, e.g. for a cylindrical expansion
along the beam axis the volume  could be interpreted as the transverse surface 
multiplied by the lifetime, $\tau$, of the system, e.g. $\pi R^2\tau$ as would be more 
appropriate for a scenario with longitudinal scaling~\cite{bjorken}.  The value of $R$ should therefore
not be considered in a full dynamical model. It is not necessarily related to the size
of the system as deduced from a HBT analysis~\cite{hbttevatron,hbtalice,hbtstar} but serves to
fix the normalization of the distribution.
The radius (volume) and the temperature do not show any significant change 
between the lowest
beam energy considered (0.54 TeV) and the highest one (7 TeV) as can be 
seen from Fig.~\ref{tsallis_T_v3}
and Fig.~\ref{tsallis_R_v3}. 
\begin{table}
\begin{center}
\begin{tabular}{|c|ccc|}
\hline
\hline
$\sqrt{s}$, TeV   &$T$, MeV         &$R$, $fm$       &$q$                \\ \hline
\hline
  0.54 (UA1)          &77.59 $\pm$ 1.40  &4.25$\pm$0.10   &1.1175 $\pm$ 0.0014 \\
  0.9  (ALICE)        &75.45 $\pm$ 3.18  &4.55$\pm$0.26   &1.1305 $\pm$ 0.0031 \\
  0.9  (ATLAS)        &83.89 $\pm$ 1.35  &3.91$\pm$0.11   &1.1217 $\pm$ 0.0007 \\
  2.36 (ATLAS)        &75.79 $\pm$ 4.01  &4.46$\pm$0.37   &1.1419 $\pm$ 0.0025 \\
  7    (ATLAS)        &82.42 $\pm$ 1.30  &4.34$\pm$0.11   &1.1479 $\pm$ 0.0008 \\ 
\hline
\hline
\end{tabular}
\caption{Parameters for the Tsallis fit}. 
\label{t1}
\end{center}
\end{table}

\begin{figure}
\begin{center}
\includegraphics[width=\textwidth,height=12cm]{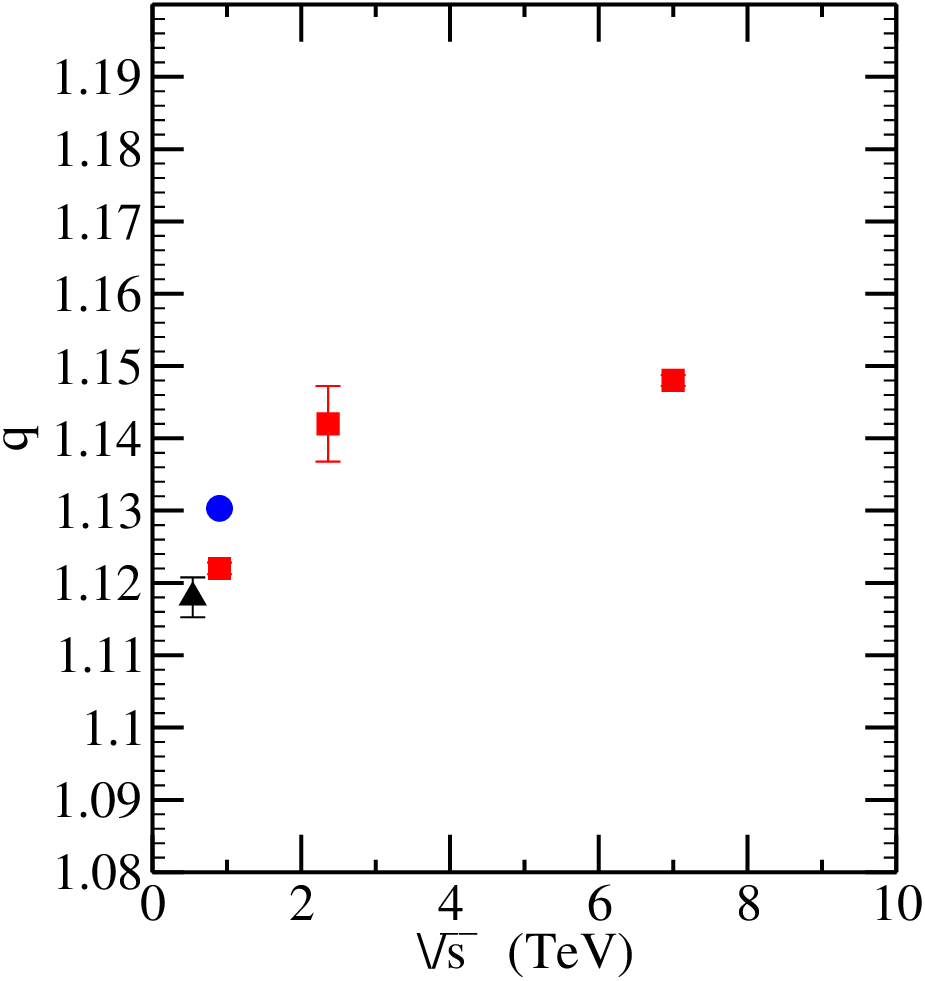}
\caption{(Color online) 
Energy dependence of the Tsallis parameter $q$ appearing in the Tsallis distribution.
Square points are from the ATLAS collaboration~\cite{ATLAS}, 
the round point is from the ALICE collaboration~\cite{ALICE} and the triangle point is from the UA1 collaboration~\cite{UA1}.
 }
\label{tsallis_q_v3}
\end{center}
\end{figure}

The energy dependence of the various parameters is displayed in 
Figs.~\ref{tsallis_T_v3},~\ref{tsallis_q_v3} and \ref{tsallis_R_v3}.

\begin{figure}
\begin{center}
\includegraphics[width=\textwidth,height=12cm]{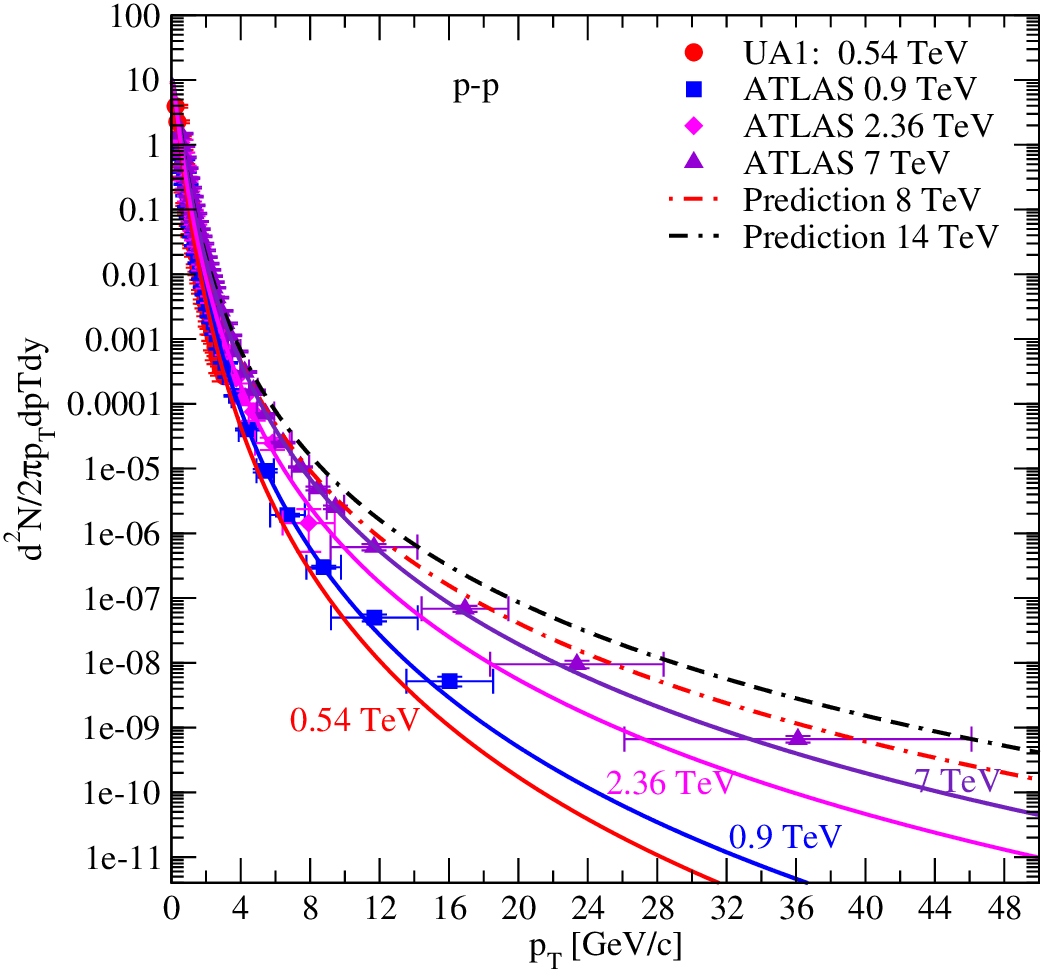}
\caption{(Color online) Charged hadron yield as a function of 
the transverse momentum $p_{T}$ from the  UA1 collaboration 
in $p\bar{p}$ collisions at $\sqrt{s}=0.54$ TeV~\cite{UA1}
and data from the ATLAS collaboration~\cite{ATLAS}. 
The curves were obtained using Eq.~\eqref{tsallisfit}. 
The parameters of the fit are given in Table~\ref{t1}. }
\label{ATLAS_fit}
\end{center}
\end{figure}

In Fig.~\ref{ATLAS_fit} we show the  charged hadron yields as a function of 
the transverse momentum $p_{T}$ at four different  energies. 
The experimental data points are from the  ATLAS collaboration~\cite{ATLAS}
for $pp$ collisions at 
energies $\sqrt{s}=0.9,~2.36,~7$ TeV  and  from the UA1 collaboration
in $p\bar{p}$ collisions at $\sqrt{s}=0.54$ TeV~\cite{UA1}. 
The curves represent fits of Eq.~\eqref{tsallisfit} to the data.

The extremely large range of $p_T$ described by Tsallis distribution
makes it applicable in the region usually considered to be the domain of
QCD hard scattering.
This may be interpreted as a manifestation of the
``duality'' between the statistical and dynamical description of strong
interactions \cite{Cleymans:2010aa,Cleymans:2011im}.
In this sense Tsallis statistics may be considered as an effective theory
allowing for an extension of the region of applicability of perturbative QCD 
from large to
low $p_T$. It is not unnatural, as approximate scale invariance manifested
in QCD both at large and small momentum scales is qualitatively similar 
to power law statistics.
It remains to be understood whether any further relations can be
found, like the dynamical origins of thermal spectra~\cite{Satz:2012aw}.
Note, that the application of the duality to the
longitudional spectra should be performed at the level
of parton distributions, where the natural temperature
scale is defined by the corresponding intrinsic transverse
momentum~\cite{Efremov:2010mt}, rather than hadrons directly,
which would lead to the ``temperature'' of beam energy scale order.

\begin{figure}
\begin{center}
\includegraphics[width=\textwidth,height=12cm]{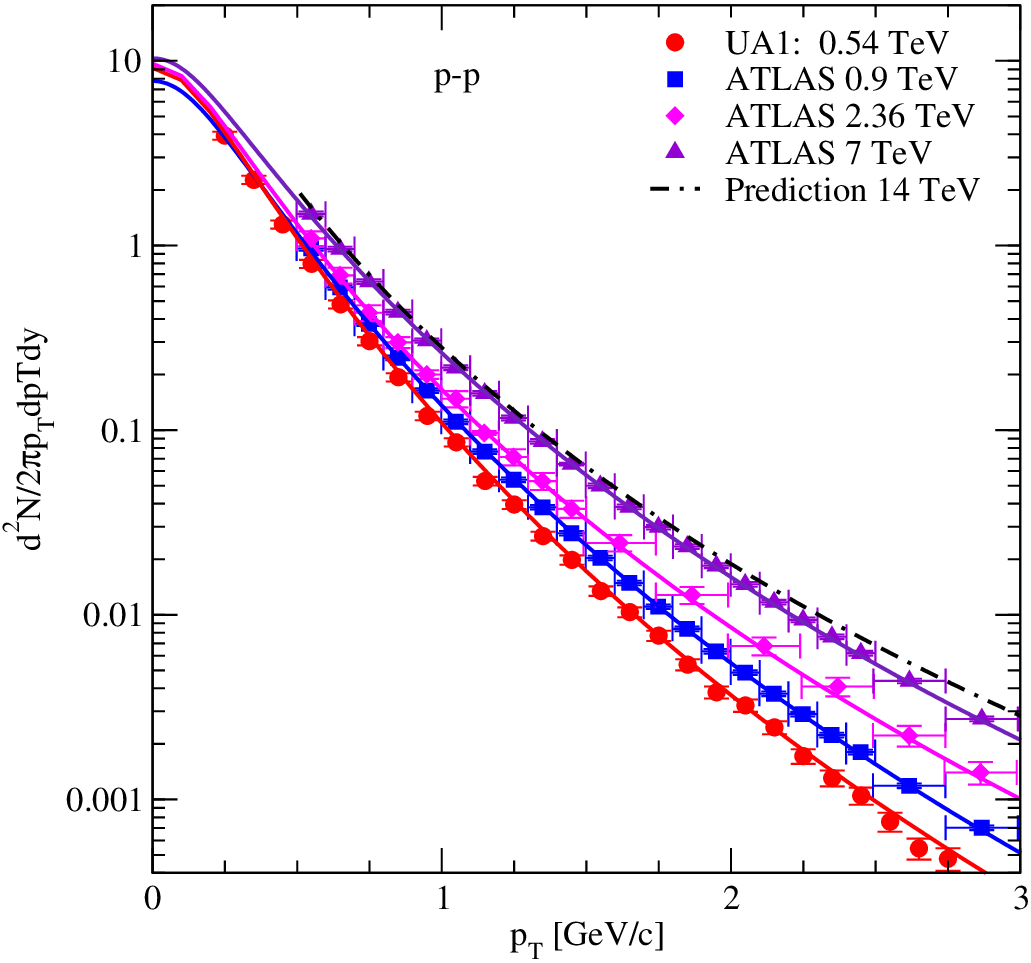}
\caption{(Color online) Charged hadron yield as a function of 
the transverse momentum $p_{T}$ from the  UA1 collaboration 
in $p\bar{p}$ collisions at $\sqrt{s}=0.54$ TeV~\cite{UA1}
and data from the ATLAS collaboration~\cite{ATLAS} showing only 
the low transverse momentum part. 
The curves were obtained using Eq.~\eqref{tsallisfit}. 
The parameters of the fit are given in Table~\ref{t1}. }
\label{ATLAS_fit_lowpt}
\end{center}
\end{figure}

The Tsallis parameter $q$ shows a clear but weak energy dependence which we have parameterized as
\begin{equation}
q(s) = 1.12567\left(\sqrt{s}\right)^{0.011211}
\end{equation}
On the basis  of these changes with energy we have attempted a prediction
for the transverse momentum distribution at invariant energy of 
8 and 14 TeV which are presented in Fig.~\ref{ATLAS_fit}.
A different parameterization has been proposed by Wibig~\cite{wibig1,wibig2}, the values obtained 
there are systematically below the values obtained here.
The fits are also very good at low transverse momenta as can be seen from 
Fig.~\ref{ATLAS_fit_lowpt}.
The curves are calculations using Eq.~\eqref{tsallisfit} at mid-rapidity $y = 0$ and zero
chemical potential $\mu = 0$ for the parameters of the Tsallis fit in Table~\ref{t1}. 
\section{Discussion and Conclusions}
As has been noticed in other publications, the Tsallis distribution 
leads to excellent fits to the transverse momentum distributions
in high energy $p-p$ collisions. 
By comparing results from UA1~\cite{UA1} to results obtained at the 
LHC~\cite{ALICE,ATLAS,CMS} it has been possible to extract the parameters
$q,T$ and $V$ for a wide range of energies.
A consistent picture emerges from a comparison of fits using the Tsallis
distribution to high energy $p-p$ collisions. 

\vspace{0.25cm}
{\bf Acknowledgments}\\
The authors are grateful to the South Africa - JINR collaboration programme
which made this research possible.
The work of  A.S.S. and O.V.T. was supported in part by the
Russian Foundation for Basic Research, Grant No. 11-02-01538-a.
\vspace{0.5cm}

{\bf References}\\

\begin{thebibliography}{90}
\bibitem{newman} M.E.J. Newman, Contemporary Physics {\textbf{46}}, 323 (2005).
\bibitem{clauset} A. Clauset, C.R. Shalizi, M.E.J. Newman, SIAM Review {\textbf{51}}, 661 (2009).
\bibitem{levy} F.~Bardou, J.-P. Bouchaud, A. Aspect, C. Cohen-Tannoudji, ``{\sl L\'evy Statistics and Laser Cooling}'', Cambridge University Press, (2002).
\bibitem{mitzenmacher} M. Mitzenmacher, Internet Mathematics, \textbf{1}, 226 (2003).
\bibitem{STAR} B. I. Abelev et al. (STAR  Collaboration), Phys. Rev. C {\textbf 75}, 064901 (2007).
\bibitem{PHENIX} A. Adare et al. (PHENIX Collaboration), Phys. Rev. C {\textbf 83}, 052004, (2010); Phys. Rev. C {\textbf{83}}, 064903 (2011).
\bibitem{ALICE} K. Aamodt, et al. (ALICE Collaboration), 
Eur. Phys. J. C {\textbf{71}} 1655 (2011); Phys. Lett. {\bf B693}, 53 (2010) ; 
Phys. Rev. D {\bf 82}, 052001 (2010).
\bibitem{ATLAS} G. Aad, et al. (ATLAS Collaboration), 
New J. Phys. {\textbf{13}}, 053033 (2011).
\bibitem{CMS} V. Khachatryan, et al. (CMS Collaboration), Phys. Rev. Lett. {\textbf 105}, 022002 (2010).
\bibitem{bediaga} I. Bediaga, E.M.F. Curado, J.M. de Miranda, Physica A {\textbf 286}, 156 (2000).
\bibitem{wilkreview} G. Wilk and Z. Wlodarczyk, Eur. Phys. J. A{\textbf 40}, 299 (2009); {\textbf 48}, 161 (2012).
\bibitem{hagedorn} R. Hagedorn, Riv. Nouvo Cimento {\textbf 6}, 1 (1984).
\bibitem{michael} C. Michael and L. Vanryckeghem, J. Phys. G {\textbf 3} L151 (1977).
%
\bibitem{tsallis1} C. Tsallis, J. Statist. Phys. {\bf 52}, 479 (1988).
\bibitem{tsallis2} C. Tsallis, R. S. Mendes, A. R. Plastino, Physica A {\bf 261}, 534 (1998).
\bibitem{miller} J. M. Conroy, H. G. Miller, A. R. Plastino, 
Phys. Lett. A {\bf 374}, 4581 (2010).
\bibitem{worku1} J. Cleymans and D. Worku, J. Phys. G {\bf 39}, 025006 (2012).
\bibitem{worku2} J. Cleymans and D. Worku, 
Eur. Phys. J. A {\bf 48}, 160 (2012). 
\bibitem{parvan1} A.S.~Parvan, Phys. Lett. A {\bf 350}, 331 (2006). 
\bibitem{parvan2} A.S.~Parvan, Phys. Lett. A {\bf 360}, 26 (2006). 
\bibitem{lavagno1} A. Lavagno, Phys. Lett. A {\textbf 301}, 13 (2002). 
\bibitem{lavagno2} A. Drago, A. Lavagno, P. Quarati, Physca A {\textbf 344}, 472 (2004). 
\bibitem{lavagno3} W.M. Alberico, A. Lavagno, Eur. Phys. J. A {\textbf 40}, 313 (2009). 
\bibitem{lavagno4} A. Lavagno, Phys. Rev. C {\textbf 81}, 044909 (2010). 
\bibitem{coalesce1} K. \"Urm\"ossy, T.S. Bir\'{o}, Phys. Lett. B {\bf 689},  14 (2010).
\bibitem{coalesce2}  K. \"Urm\"ossy, T.S. Bir\'{o}, J. Phys. G {\bf 36},  064044 (2009).
%
%
\bibitem{wong} Cheuk-Yin Wong, G. Wilk, Acta Physica Polonica, {\bf 43}, 2047 (2012).
%
\bibitem{deppman} L. Marques, E.Andrade-II, A. Deppman, arXiv:1210.1725[hep-ph]
\bibitem{deppman1} A. Deppman, arXiv:1212.0379[hep-ph].
%
%
\bibitem{deppman2} I. Sena, A. Deppman, arXiv:1208.2952[hep-ph]
\bibitem{deppman3} I. Sena, A. Deppman, arXiv:1209.2367[hep-ph]
%
\bibitem{urmossy} K. \"Urm\"ossy, arXiv:1212.0260[hep-ph].
%
\bibitem{gena} A. Bednyakov, A.A. Grinyuk, G.I. Lykasov, M.~Poghosian,  
Int. J.  Mod. Phys. A {\bf  27}, 1250042 (2012).
\bibitem{UA1} C.~Albajar et al. (UA1 Collaboration), Nucl. Phys. B {\bf 335} (1990) 261.
\bibitem{bjorken} J.D. Bjorken, Phys. Rev. D {\bf 27} (1983) 140.
\bibitem{hbttevatron} T. Alexopoulos, Phys. Rev. D {\bf 48} (1993) 1931.
\bibitem{hbtalice} K. Aamodt wt al., Phys. Rev. D {\bf 82} (2010) 052001.
\bibitem{hbtstar} M.M. Aggarwal, Phys. Rev. V {\bf 83} (2011) 064905.
\bibitem{Cleymans:2010aa}
  J.~Cleymans, G.I.~Lykasov, A.N.~Sissakian, A.S.~Sorin and
O.V.~Teryaev,
arXiv:1004.2770[hep-ph].
\bibitem{Cleymans:2011im}
  J.~Cleymans, G.I.~Lykasov, A.S.~Sorin and O.V.~Teryaev,
Phys.\ Atom.\ Nucl.\  {\bf 75}, 725 (2012)
[arXiv:1104.0620 [hep-ph]].  

\bibitem{Satz:2012aw}
  H.~Satz, Int.\ J.\ Mod.\ Phys.\ E {\bf 21}, 1230006 (2012).
\bibitem{Efremov:2010mt}
A.~V.~Efremov, P.~Schweitzer, O.~V.~Teryaev and P.~Zavada,
Phys.\ Rev.\ D {\bf 83}, 054025 (2011)
[arXiv:1012.5296 [hep-ph]].
\bibitem{wibig1} T. Wibig,  J. Phys. G: Nucl. Part. Phys. {\textbf 37} 115009 (2010).
\bibitem{wibig2} T. Wibig, I. Kurp, JHEP {\textbf 0312} 039 (2003).
\end{thebibliography}
\end{document}